\documentclass[]{pasj01}
\Received{2019/12/06}
\Accepted{2019/12/27}
\begin{document} 

\title{ 
%\LETTERLABEL %%% <-- uncomment for LETTER article  
%\REVIEWLABEL %%% <-- uncomment for REVIEW article  

First Detection of X-Ray Line Emission from Type IIn Supernova 1978K with \textit{XMM-Newton}'s RGS}

%%% begin:list of authors
% Do NOT capitalize all letters in "textsc".
\author{Yuki \textsc{Chiba}\altaffilmark{1,}$^{*}$}%
%\thanks{Example: Present Address is xxxxxxxxxx}}
\altaffiltext{1}{Graduate School of Science and Engineering, Saitama University, 255 Shimo-Okubo, Sakura, Saitama, Saitama 338-8570, Japan }
\email{chiba@heal.phy.saitama-u.ac.jp}

\author{Satoru \textsc{Katsuda}\altaffilmark{1,}$^{*}$}
\email{katsuda@phy.saitama-u.ac.jp}

\author{Takashi \textsc{Yoshida}\altaffilmark{2,}$^{*}$}
\altaffiltext{2}{Department of Astronomy, Graduate School of Science, the University of Tokyo, Tokyo, 113-0033, Japan}
\email{tyoshida@astron.s.u-tokyo.ac.jp}

\author{Koh \textsc{Takahashi}\altaffilmark{3,}$^{*}$}
\altaffiltext{3}{Max Plank Institute for Gravitational Physics (Albert Einstein Institute), D-14476 Potsdam, Germany}
\email{koh.takahashi@aei.mpg.de}

\author{Hideyuki \textsc{Umeda}\altaffilmark{2,}$^{*}$}
\email{umeda@astron.s.u-tokyo.ac.jp}

%%% end:list of authors

%% `\KeyWords{}' always has to be placed before ``\maketitle'' 
%%  List of Key Words:  https://academic.oup.com/pasj/pages/Pasj_Keywords 

\maketitle

\begin{abstract}

We report on robust measurements of elemental abundances of the Type IIn supernova SN~1978K, based on the high-resolution X-ray spectrum obtained with the Reflection Grating Spectrometer (RGS) onboard \textit{XMM-Newton}.  The RGS clearly resolves a number of emission lines, including N Ly$\alpha$, O Ly$\alpha$, O Ly$\beta$, Fe XVII, Fe XVIII, Ne He$\alpha$ and Ne Ly$\alpha$ for the first time from SN~1978K.  The X-ray spectrum can be represented by an absorbed, two-temperature thermal emission model, with temperatures of $kT \sim 0.6$\,keV and $2.7$\,keV.  The elemental abundances are obtained to be N $=$ $2.36_{-0.80}^{+0.88}$, O $=$ $0.20 \pm{0.05}$, Ne $=$ $0.47 \pm{0.12}$, Fe $=$ $0.15_{-0.02}^{+0.01}$ times the solar values.  The low metal abundances except for N show that the X-ray emitting plasma originates from the circumstellar medium blown by the progenitor star.  The abundances of N and O are far from CNO-equilibrium abundances expected for the surface composition of a luminous blue variable, and resemble the H-rich envelope of less-massive stars with masses of 10--25\,M$_\odot$.  Together with other peculiar properties of SN~1978K, i.e., a low expansion velocity of 500--1000\,km\,s$^{-1}$ and SN~IIn-like optical spectra, we propose that SN~1978K is a result of either an electron-capture SN from a super asymptotic giant branch star, or a weak Fe core-collapse explosion of a relatively low-mass ($\sim$10\,M$_\odot$) or high-mass ($\sim$20--25\,M$_\odot$) red supergiant star.  However, these scenarios can not naturally explain the high mass-loss rate of the order of $\dot{M} \sim 10^{-3}\,\rm{M_{\odot}\ yr^{-1}}$ over $\gtrsim$1000\,yr before the explosion, which is inferred by this work as well as many other earlier studies.  Further theoretical studies are required to explain the high mass-loss rates at the final evolutionary stages of massive stars.

\end{abstract}

\section{Introduction}

Massive stars end their lives in supernova (SN) explosions, but the very last evolutionary stage of a massive star is still debated.  Even if progenitor stars happen to be detected in pre-explosion images, such data are usually too poor to determine the progenitor properties.  Thus, an important clue to investigating progenitors is the circumstellar medium (CSM) blown by the progenitors during their final evolutionary stages.  When the CSM is heated by the interaction between the rapidly expanding SN ejecta and the slowly moving CSM, we are able to study properties of the CSM and the progenitor through multiwavelength (radio, infrared, ultraviolet, and X-ray) observations.  Among various subtypes of SNe (IIP, IIL, IIb, Ib, Ic, etc.), the IIn, defined broadly as SNe that exhibit bright and narrow Balmer lines of H in their spectra (e.g., \cite{schlegel1990}), exhibit the strongest ejecta-CSM interactions, providing us with the best opportunities to study the progenitor and its later evolution through the observed properties of the CSM (e.g., \cite{moriya2013,smith2017,chandra2018}).  

A large number of X-ray observations of ejecta-CSM interactions have been performed in the last two decades (e.g., \cite{dwarkadas2012,chandra2018}).  Most of these studies are based on light curves, from which we can reveal density structures of the CSM, namely mass-loss histories of the progenitors.  The mass-loss rates have been used to constrain the initial masses of the progenitor stars; e.g., $\dot{M} \sim\ 10^{-6}$\,M$_\odot$\ $\rm{yr^{-1}}$ indicates supergiants (e.g., \cite{maeda2014}), whereas $\dot{M} \gtrsim 10^{-3}$\,M$_\odot$\ $\rm{yr^{-1}}$ indicates luminous blue variables (LBVs) (e.g., \cite{katsuda2014}).  In addition, recent X-ray studies have been revealing spectral softening with time possibly caused by decreasing absorbing materials as a shock overtakes the absorbing CSM (e.g., \cite{chandra2012}), and such a spectral evolution may indicate a torus-like geometry of the CSM \citep{katsuda2016}.

In addition to the luminosity and spectral evolution, the chemical composition of the CSM should be another important, but often missing, clue to the progenitor stars.  In particular, the relative abundances of C, N, and O elements give an indication of the degree of CNO-processing in the surface layers, and allow us to infer the evolutionary state of the progenitor star.  However, there are only five SNe (SNe~1979C, 1987A, 1993J, 1995N, and 1998S) for which CNO abundances in the shocked CSM were measured, as summarized in \citet{fransson2005cno}.  In all cases, strong N enhancements were found, evidencing CNO processing to some degrees.

SN~1978K is an X-ray--luminous SN located at the outskirt of the nearby (4.61\,Mpc: \cite{qing2015distance}) galaxy NGC~1313. The X-ray emission of SN~1978K was first uncovered with the \textit{ROSAT} Position Sensitive Proportional Counter in 1992 \citep{ryder1993sn}. \citet{ryder1993sn} also suggested that SN~1978K was a Type II SN that exploded around 1978 May 22 \citep{montes1997}.  After that, \citet{schlegel1999physical} suggested that the nature of SN~1978K found from the X-ray and radio data is typical of SNe IIn.

Notably, SN~1978K remains bright almost 40 years after the explosion (e.g., \cite{lenz2007silver,smith2007multiwavelength,zhao2017x,smith2019}).  The mass-loss rates estimated in radio, H$\alpha$, infrared, and X-rays range from $10^{-4}\,\rm{M_{\odot}\ yr^{-1}}$ to $10^{-2}\,\rm{M_{\odot}\ yr^{-1}}$ (e.g., \cite{ryder1993sn,chugai1995massloss,schlegel1999physical,tanaka2012infrared,kuncarayakti2016evolving}).  Such a high mass-loss rate suggests that the progenitor is an LBV.  Meanwhile, a possible progenitor of SN~1978K was identified in pre-explosion images \citep{ryder1993sn}.  The B-band magnitude was estimated to be $B_{\rm{J}} = 22.1$ mag in 1974--1975, which corresponds to an absolute magnitude of $-6$ mag, assuming no reddening.  This is somewhat dark as an LBV (mass $>$ 35\,M$_\odot$), but is typical as a supergiant (mass 10--25\,M$_\odot$).  Also, age-dating of the stellar population surrounding SN~1978K, suggested a relatively small progenitor mass of 8.8$\pm$0.2\,M$_\odot$ \citep{Williams2018mass}.

Despite long-term X-ray follow-up observations of SN~1978K with various X-ray observatories such as \textit{ASCA}, \textit{XMM-Newton}, and \textit{Chandra}, little is known about elemental abundances.  \citet{petre1994} analyzed \textit{ASCA} PV phase observations of SN~1978K, finding featureless X-ray spectrum.  They reported that the X-ray spectrum is described by either a powerlaw with a photon index $\Gamma \sim 2.2$ or a thermal model with a temperature $kT \sim 3$\,keV and abundances $Z \sim 0.2 Z_\odot$.  \citet{schlegel2004chandra} analyzed observations with \textit{Chandra} in 2002 and \textit{XMM-Newton} in 2000, and reported that the spectra were best fitted by a two-temperature, variable-abundance, optically thin gas model with temperatures of 0.6\,keV and 3\,keV.  They also argued an elevated Si abundance, with Si/H $=$ $3.20_{-1.90}^{+1.80}$ (Si/H)$_\odot$ for the soft component.  However, a later study by \citet{lenz2007silver}, who analyzed \textit{Chandra} and \textit{XMM-Newton} observations in 2003, did not confirm such a Si enhancement.  More recently, \citet{smith2007multiwavelength} analyzed \textit{XMM-Newton} data in 2000--2006, and interpreted the featureless X-ray spectrum as being due to a large abundance of He.  They speculated that the shock is located in a He-rich layer that was ejected by the progenitor star or the SN explosion.  However, we should be cautious about this interpretation, given that there are no lines from H and He in the X-ray spectrum, and thus it is generally difficult for any X-ray instruments to measure the He/H ratio.

The region including SN~1978K has been visited by {\it XMM-Newton} a number of times to monitor not only SN~1978K but also another two ultra luminous X-ray sources (X-1 and X-2) in NGC~1313.  These data were recently used to study long-term X-ray light curves of SN~1978K. In this paper, we also make use of the rich {\it XMM-Newton} data, but mainly focus on the data obtained with the Reflection Grating Spectrometer (RGS; \cite{denherder2001}) to measure elemental abundances in precisely details.  

The paper proceeds with Section 2 describing the observations and data reduction, Section 3 with the analysis and results, Section 4 with discussions, and ends with the summary. Throughout, we adopt a distance to NGC~1313 of 4.61\,Mpc, as determined by \citet{qing2015distance}. This is slightly farther compared to the values of 4.5\,Mpc used in \citet{ryder1993sn}, and 4.13\,Mpc in \citet{schlegel2004chandra}, \citet{smith2007multiwavelength} and \citet{lenz2007silver}.

\begin{table*}[htbp]
  \tbl{\textit{XMM-Newton} observations analyzed in this paper.}{%:
  \begin{tabular}{ccccccc}
      \hline
      No. & Observation ID & Date & Target & Exposure & \multicolumn{2}{c}{Effective Exposure} \\ 
            &                             & (UT)  &              &  (ks)     &  \multicolumn{1}{c}{RGS1/2}(ks) & \multicolumn{1}{c}{MOS1/2 (ks)}   \\
	 \hline
    1 & 0301860101 & 2006 Mar 06 & SN 1978K  & 21.8 & \multicolumn{1}{c}{21.3/21.3} & \multicolumn{1}{c}{21.3/21.3} \\
    2 & 0693850501 & 2012 Dec 16 & NGC 1313 & 125.2 & \multicolumn{1}{c}{113.0/113.0} & \multicolumn{1}{c}{123.6/123.7} \\
    3 & 0693851201 & 2012 Dec 22 & NGC 1313  & 125.2 & \multicolumn{1}{c}{91.5/91.7} & \multicolumn{1}{c}{123.6/123.7} \\
    4 & 0722650101 & 2013 Jun 08 & SN 1978K   & 30.7 & \multicolumn{1}{c}{30.6/30.5} & \multicolumn{1}{c}{30.1/30.2} \\
    5 & 0742490101 & 2015 Mar 30 & SN 1978K   & 103.0 & \multicolumn{1}{c}{96.1/96.2} & \multicolumn{1}{c}{98.0/100.7} \\
    6 & 0764770401 & 2016 Mar 23 & NGC 1313 X-2 & 33.0 & \multicolumn{1}{c}{27.9/27.9} & \multicolumn{1}{c}{31.3/31.2} \\
    7 & 0782310101 & 2016 Oct 08 & SN 1978K    & 91.0 & \multicolumn{1}{c}{89.6/89.6} & \multicolumn{1}{c}{88.4/88.4} \\
    8 & 0803990201 & 2017 Jun 20 &NGC 1313 X-1&133.8 &\multicolumn{1}{c}{130.7/130.9}&\multicolumn{1}{c}{130.5/130.7} \\
    9 &0803990501 & 2017 Dec 07 & NGC 1313 X-1 & 128.9 & \multicolumn{1}{c}{51.3/51.4} & \multicolumn{1}{c}{125.8/125.8} \\
   10 & 0803990601&2017 Dec 09 &NGC 1313 X-1 & 128.9 & \multicolumn{1}{c}{67.1/67.2}  & \multicolumn{1}{c}{125.8/125.9} \\
	 \hline
  \end{tabular}}\label{tab:obs_info}
\begin{tabnote}
%\footnotemark[$*$] Total clean exposure times of RGS. \\ 
%\footnotemark[$\dag$] Explanation of value 3. 
%\footnotemark[$\ddag$]  ... \\ 
%\footnotemark[$\S$]  ... \\ 
%\footnotemark[$\|$]  ... \\
%\footnotemark[$\sharp$]  ... \\  
%\footnotemark[$**$]  ... \\ 
%\footnotemark[$\dag\dag$]  ... \\ 
\end{tabnote}
\end{table*}

\section{Observations and data reduction}

NGC~1313 has been observed with \textit{XMM–Newton} many times in a time interval of 13 years, from 2004 August 23 to 2017 December 09, resulting in 13 observations that cover SN~1978K.  Of these, two data sets taken in 2004 August and 2014 July 06 were excluded because of the high soft-proton background and the overlapping of NGC 1313 X-1 and SN~1978K along the dispersion direction of the RGS, respectively.  The data on 2017 June 14 was also excluded, because SN~1978K was detected at the edge of the field of view of the RGS.  As a result, we adopted 10 observations in this work.  We further excluded time periods suffering from soft-proton backgrounds following the Science Analysis System (SAS) standard procedures.  Thus-derived cleaned exposure times, i.e., effective exposure times, as well as other detailed information of these observations are summarized in Table~\ref{tab:obs_info}.  

\begin{figure*}[htbp]
 \begin{center}
   \includegraphics[width=12.5cm]{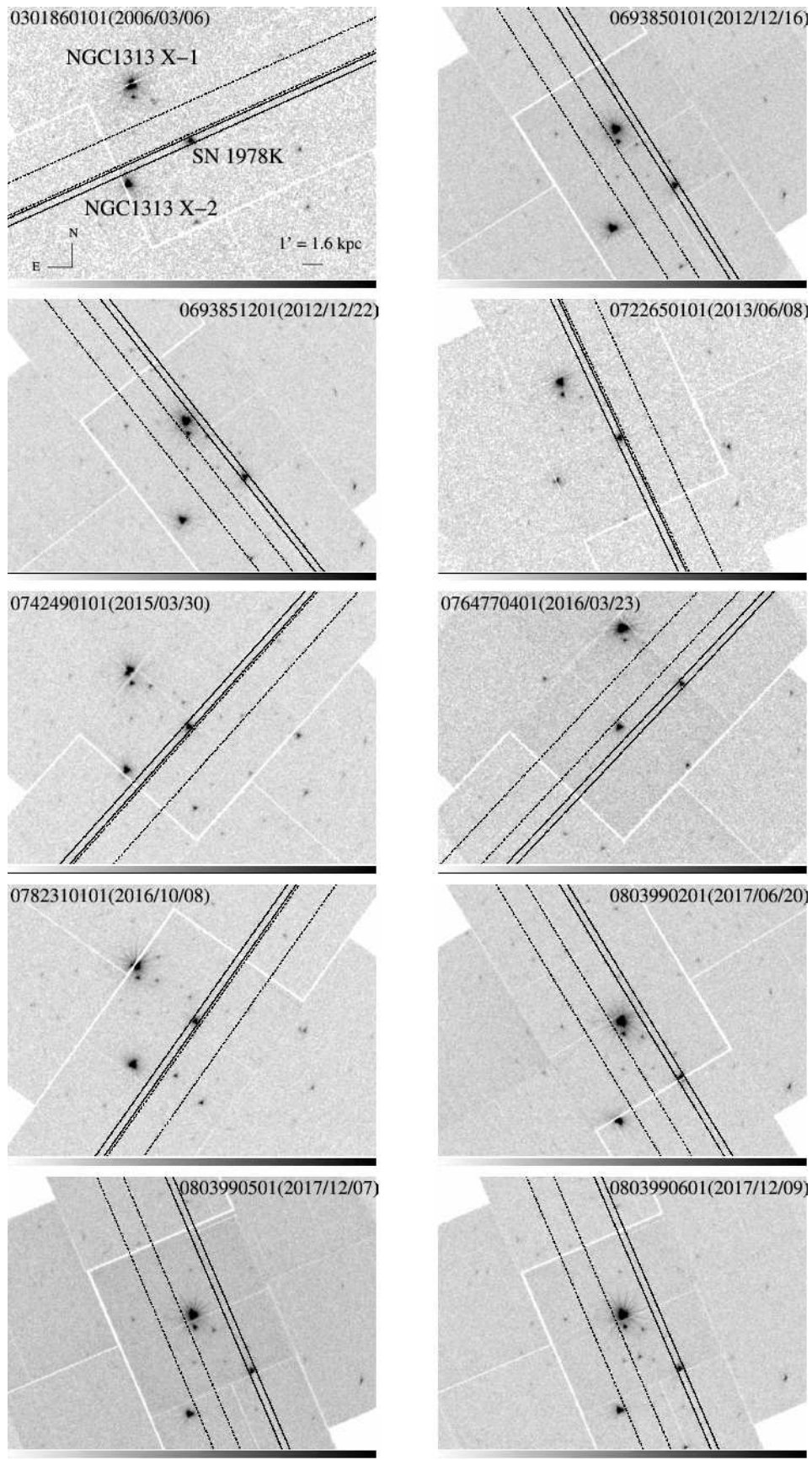} 
 \end{center}
\caption{EPIC MOS images of NGC~1313, with the intensity scale of square root.  The strips enclosed within solid and dashed lines are the RGS source and background extraction region, respectively. }\label{fig:extra_srcbgd_reg}
\end{figure*}

All the raw data were reduced with the \textit{XMM-Newton} SAS v16.1.0.  For each observation, we extracted the first-order RGS spectra from a \timeform{30''}-width cross-dispersion region, using the \textit{rgsproc} task with the option spectrumbinning=lambda.  The net integration time after cleaning was $\sim$720\,ks, and we combined RGS1 and RGS2 in all the ten observations into one data file with the task \textit{rgscombine} to improve the photon statistics.  We extracted background spectra from blank fields in the same observations, as indicated in Figure \ref{fig:extra_srcbgd_reg}.  

We also analyzed data obtained with the EPIC (European Photon Imaging Camera) MOS \citep{turner2001}.  We extracted spectra from a circular region of \timeform{30''} diameter centered on SN~1978K.  The background spectra were extracted from the surrounding annular region with a radius of \timeform{42.4''}.  \\

\section{Analysis and results}

\begin{figure*}[htbp]
 \begin{center}
  \includegraphics[width=17cm]{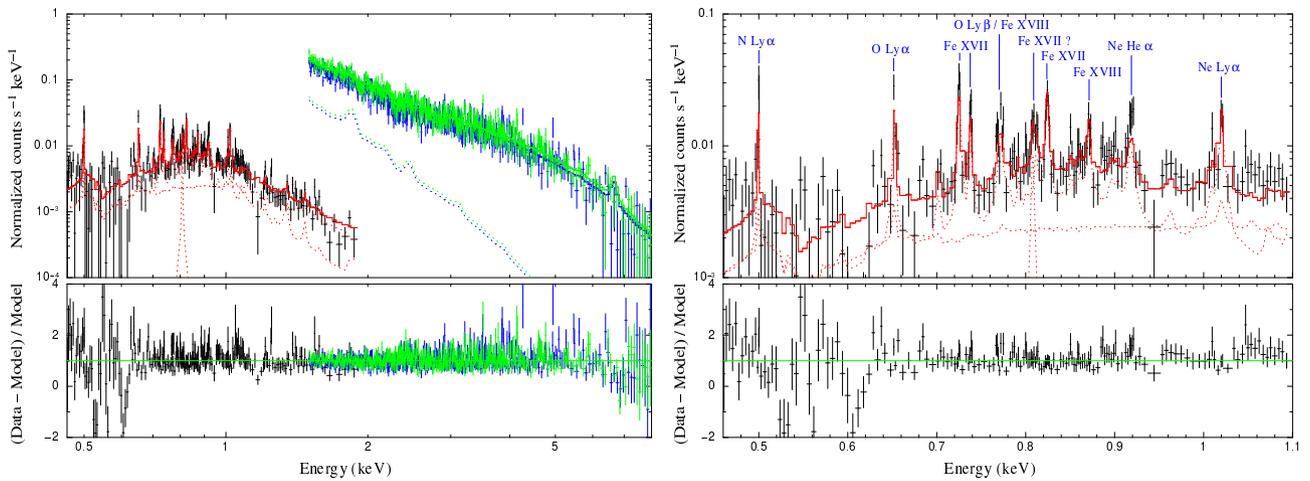} 
 \end{center}
\caption{Left: Combined RGS1+RGS2, MOS1, MOS2 spectra in black, blue, and green, respectively.  The best-fit model consisting of two-component VAPEC plus a Gaussian at 0.809\,keV is also shown.  The lower panel shows ratios of data to the best-fit model.  Right: Same as left but focuses on the RGS data.  Prominent lines are identified.}\label{fig:spec}
\end{figure*}

We present the RGS and MOS spectra in Figure~\ref{fig:spec}.  The RGS spectrum clearly resolves a number of emission lines, including N Ly$\alpha$, O Ly$\alpha$, O Ly$\beta$, Fe XVII, Fe XVIII, Ne He$\alpha$, and Ne Ly$\alpha$ for the first time from SN~1978K.  We fitted the MOS and RGS spectra simultaneously with a thermal emission model, using XSPEC version 12.10.0 \citep{arnaud1996xspec}.  For our spectral modeling, we followed previous X-ray spectral analyses (e.g., \cite{schlegel2004chandra,lenz2007silver,zhao2017x}), adopting an absorbed (TBabs: \cite{wilms2000}), two-temperature, variable-abundance thermal emission model in ionization equilibrium, i.e., the VAPEC model in XSPEC.  We restricted the RGS data in 0.46--1.9\,keV where we obtained good signal-to-noise ratios, and MOS data in 1.5--8.0\,keV where the MOS data have much higher signal-to-noise ratios than the RGS.  To find the best-fit parameters and estimate their uncertainties, we used \textit{C}-statistics (\cite{cash1979parameter}).  

The VAPEC model allows us to vary abundances of any astrophysically important elements.  Here, we treat abundances of N, O, Ne, Mg,  Al, Si, S, Ar, Ca, Fe and Ni as free parameters, because K- and/or L-shell emission lines from these elements fall in the energy range of our X-ray spectra.  As described above, we can see clear line emission from N, O, Ne, and Fe in our RGS spectrum (Figure~\ref{fig:spec}), which assures that we can measure their abundances.  However, other elements, i.e., Mg, Al, Si, S, Ar, Ca, and Ni show little or no evidence for line emission.  Abundances of these elements are tied together with those of abundant elements that are nearby in atomic number.  Details of the treatment can be found in Table~\ref{tab:spec_fitpara}. \citet{smith2007multiwavelength} set the abundance of He as a free parameter.  However, it is in principle difficult to measure the He abundance, as we described in Section~1.  Therefore, we assumed the He abundance to be 2.6 times the solar value \citep{angr1989}, which is the same as SN~1987A \citep{lundqvist1995line} whose optical spectrum including He line emission is very similar to SN~1978K \citep{kuncarayakti2016evolving}.  The abundances are tied together between the two VAPEC components.   
We fix the redshift parameter to the known value for NGC~1313 \citep{kobulnicky2004metallicities} and allow the line broadening to vary freely by using gsmooth in XSPEC.  We assume the width to be proportional to the line energies.

After fitting, we found an excessing residual structure around 0.81\,keV.  By adding a Gaussian, we measured the line centroid to be $0.809 \pm{0.001}$ keV after correcting for the redshift.  This is not consistent with O Ly$\gamma$ at 0.817\,keV nor the well-known intense Fe XVII 2p--3d lines at 0.812\,keV and 0.826\,keV (e.g., \cite{gillaspy2011fe}).  It may arise from a missing (or stronger-than-expected) Fe/Ni L-shell line.  However, we should caution that this line is not significant in the spectra in two long-exposure data, No.\ 7 and 8 in Table \ref{tab:obs_info}.  Therefore, at this moment, it is unclear whether it is a true emission line or not.  In this context, we add a Gaussian component to simply improve the fit quality in the following fitting procedure.

As shown in Figure~\ref{fig:spec}, this model successfully fits our data.  The fit results are summarized in Table~\ref{tab:spec_fitpara}.  The best-fit low and high temperatures of $kT_{\rm{l}}$ $=$ $0.63 \pm{0.02}$ keV and $kT_{\rm{h}}$ $=$ $2.74_{-0.05}^{+0.06}$ keV are consistent with those obtained with \textit{XMM-Newton} data \citep{lenz2007silver}.  On the other hand, the absorbing H column density of $N_{\rm{H}}$ $=$ $0.16 \pm{0.01}$ $\times$ $10^{22} \ \rm{cm}^{-2}$ is reduced by $\sim$50\% compared with the most recent X-ray study \citep{zhao2017x}.  This is most likely due to the difference of abundance tables between the previous study using a modern table by \citet{wilms2000} and our work using a classic table \citep{angr1989} which has a higher metallicity by a factor of a few 10\%.  

\begin{figure}[htbp]
 \begin{center}
  \includegraphics[width=8.5 cm]{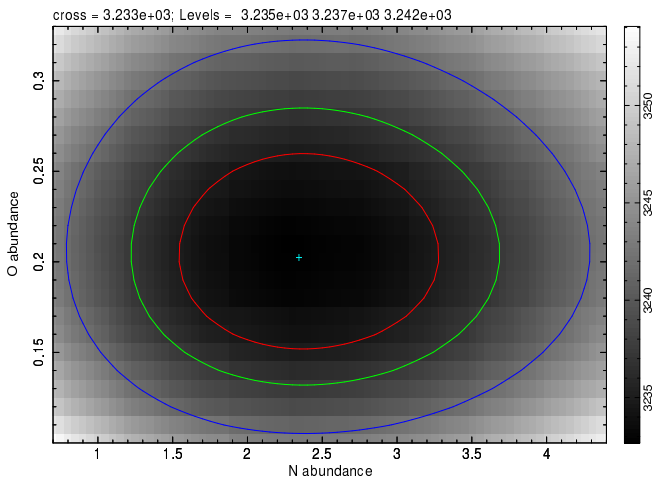} 
 \end{center}
\caption{Confidence contours of the N abundance vs.\ the O abundance.  Red, green, and blue lines represent 68\%, 90\% and 99\% confidence levels, respectively.} \label{fig:N_O_cont}
\end{figure}

The elemental abundances were obtained to be N $=$ $2.36_{-0.80}^{+0.88}$, O $=$ $0.20 \pm{0.05}$, Ne $=$ $0.47 \pm{0.12}$, Mg $=$ $0.25 \pm{0.04}$, Fe $=$ $0.15_{-0.02}^{+0.01}$ times the solar values.  This is the first time that a robust abundance measurement based on line emission has been achieved for SN~1978K.  Remarkably, we found an elevated N abundance.  To derive a conservative statistical uncertainty on the N abundance, we investigated a confidence contour between N and O abundances, as shown in Figure \ref{fig:N_O_cont}.  Based on this plot, we estimate N/O $\sim 12_{-6}^{+8}$ (N/O)$_\odot$ at a 90\% confidence level.  

We obtained a line broadening (one sigma) at 6\,keV to be $\lesssim$5\,eV.  This is equivalent to $\lesssim$0.54\,eV at O Ly$\alpha$, giving $kT_{\rm O} \lesssim$ 10\,keV.  If we assume collisionless-shock heating as is usually the case for SN remnants, then we can constrain the shock speed of $\lesssim$570\,km\,s$^{-1}$.  This is consistent with both the velocity measured from optical line widths (e.g., \cite{kuncarayakti2016evolving}) and the upper limit of average expansion velocity constrained by VLBI observations \citep{ryder2016}.

\begin{figure}[htbp]
 \begin{center}
  \includegraphics[width=8.9 cm]{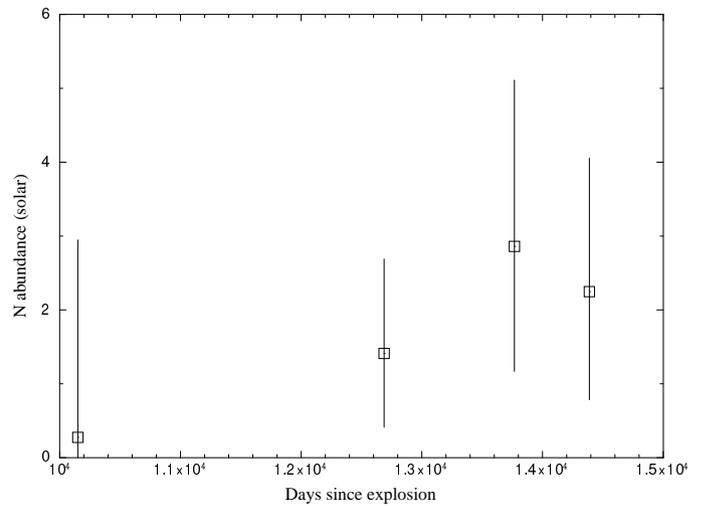} 
 \end{center}
\caption{N abundance as a function of days since explosion on 1978 May 22.} \label{fig:N_day}
\end{figure}

We then examined the time variability of the N abundance. We combined data whose observation dates are close with each other to improve photon statistics.  Specifically, we combined data taken from 2012 December 16 to 2013 June 08 (No.\ 2, 3 and 4 in Table \ref{tab:obs_info}), from 2015 March 30 to 2016 October 08 (No.\ 5, 6 and 7), and from 2017 June 20 to 2017 December 09 (No.\ 8, 9 and 10).  We then fitted the combined data with the same plasma model as we described above.  The best-fit N abundances were shown in Figure \ref{fig:N_day}, from which no significant change was found.    

\begin{figure}[htbp]
 \begin{center}
  \includegraphics[width=8.5 cm]{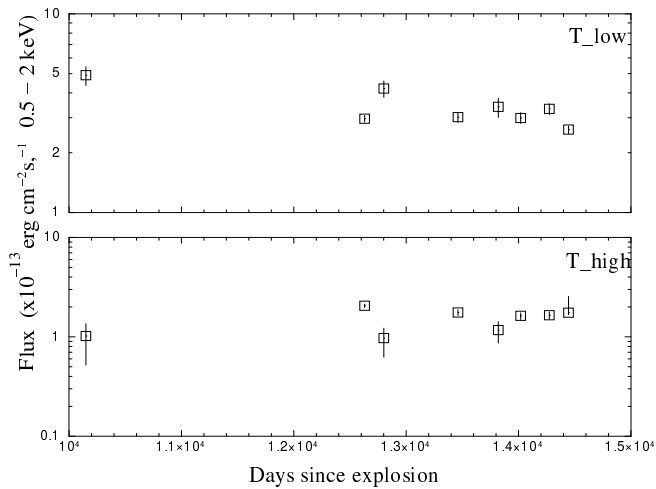} 
 \end{center}
\caption{Unabsorbed X-ray fluxes in 0.5--2\,keV for the low (top) and high (bottom) temperature components.}\label{fig:flux}
\end{figure}

We also examined luminosity variations for the low- and high-temperature components.  To this end, we used only MOS spectra, because they are much richer in photon statistics than the RGS spectra.  We again combined data that were taken closely in time, resulting in eight observation epochs.  We adopted the same model as we did for the stacked RGS plus MOS spectra, with all parameters except for the normalizations held fixed to the values in Table~\ref{tab:spec_fitpara}.  Figure~\ref{fig:flux} shows the best-fit luminosities in 0.5--2\,keV corrected for the absorption.  The luminosities for both of the two components stay constant with time.  This conflicts with a recent argument by \citet{zhao2017x} that the X-ray light curves in 0.5--2\,keV and 2--10\,keV decline as $t^{-1}$ from 2000 to 2015.  The reason of this discrepancy is unclear.  However, it is worth noting that the unabsorbed (observed) 0.5--2\,keV flux in \citet{zhao2017x} is nearly constant, whereas the 2--8\,keV flux decreases significantly.  The soft-band flux is heavily affected by the intervening absorption, which was left as a free parameter in \citet{zhao2017x}, but is fixed in this work.  The different treatment of the absorption might result in the different behavior in the soft X-ray light curve, leaving the 0.5--2\,eV flux decline argued by \citet{zhao2017x} ambiguous.

\begin{table*}[htbp]
  \tbl{Best-fit spectral parameters.}{%:
  \begin{tabular}{ccccccccc}
      \hline
       \multicolumn{6}{l}{VAPEC} & \multicolumn{3}{l}{Gaussian} \\
      \hline
      $N_{\rm{H}}$ & $kT_{\rm{l}}$ & $kT_{\rm{h}}$ & EM$_{\rm{l}}$ & EM$_{\rm{h}}$ & Redshift & LineE & Sigma & Norm \\ 
      ($10^{22}\,\rm{cm}^{-2}$) & (keV) & (keV) & ($10^{61}\,\rm{cm}^{-3}$) & ($10^{61}\,\rm{cm}^{-3}$) & ($10^{-3}$) & (keV) & (keV) & ($10^{-6}\,\rm{photons/{cm}^{-2}/s}$) \\
      \hline
      $0.16 \pm{0.01}$ & $0.63 \pm{0.02}$ & $2.74_{-0.05}^{+0.06}$ & $6.05 \pm{0.33}$ & $6.15 \pm{0.10}$ & 1.57 (Fixed)  &  $0.809 \pm{0.001}$ & 0 (Fixed) & $1.41_{-0.44}^{+0.92}$  \\
      \hline
      \multicolumn{6}{l}{Abundance} &\multicolumn{2}{l}{gsmooth} \\
      \hline
      He & N & O(=C) & Ne & Mg=Al=Si=S=Ar=Ca &  Fe=Ni  & Sigma & Index & $C$/dof \\
      (solar) &&&&&& (at 6\,keV) & \\
      \hline
      2.60 (Fixed) & $2.36_{-0.80}^{+0.88}$ & $0.20 \pm{0.05}$ & $0.47 \pm{0.12}$ & $0.25 \pm{0.04}$ & $0.15_{-0.02}^{+0.01}$  & $<$0.005 & 1 (Fixed) & 3233/3008\\
      \hline
  \end{tabular}}\label{tab:spec_fitpara}
\begin{tabnote}
\footnotemark Note- Errors represent 90\% confidence levels.  Abundances are relative to the solar values \citep{angr1989}. \\
%\footnotemark[$*$] ... \\ 
%\footnotemark[$\dag$] ... \\
%\footnotemark[$\ddag$]  ... \\ 
%\footnotemark[$\S$]  ... \\ 
%\footnotemark[$\|$]  ... \\
%\footnotemark[$\sharp$]  ... \\  
%\footnotemark[$**$]  ... \\ 
%\footnotemark[$\dag\dag$]  ... \\ 
\end{tabnote}
\end{table*}

\begin{figure}[htbp]
 \begin{center}
  \includegraphics[width=8 cm]{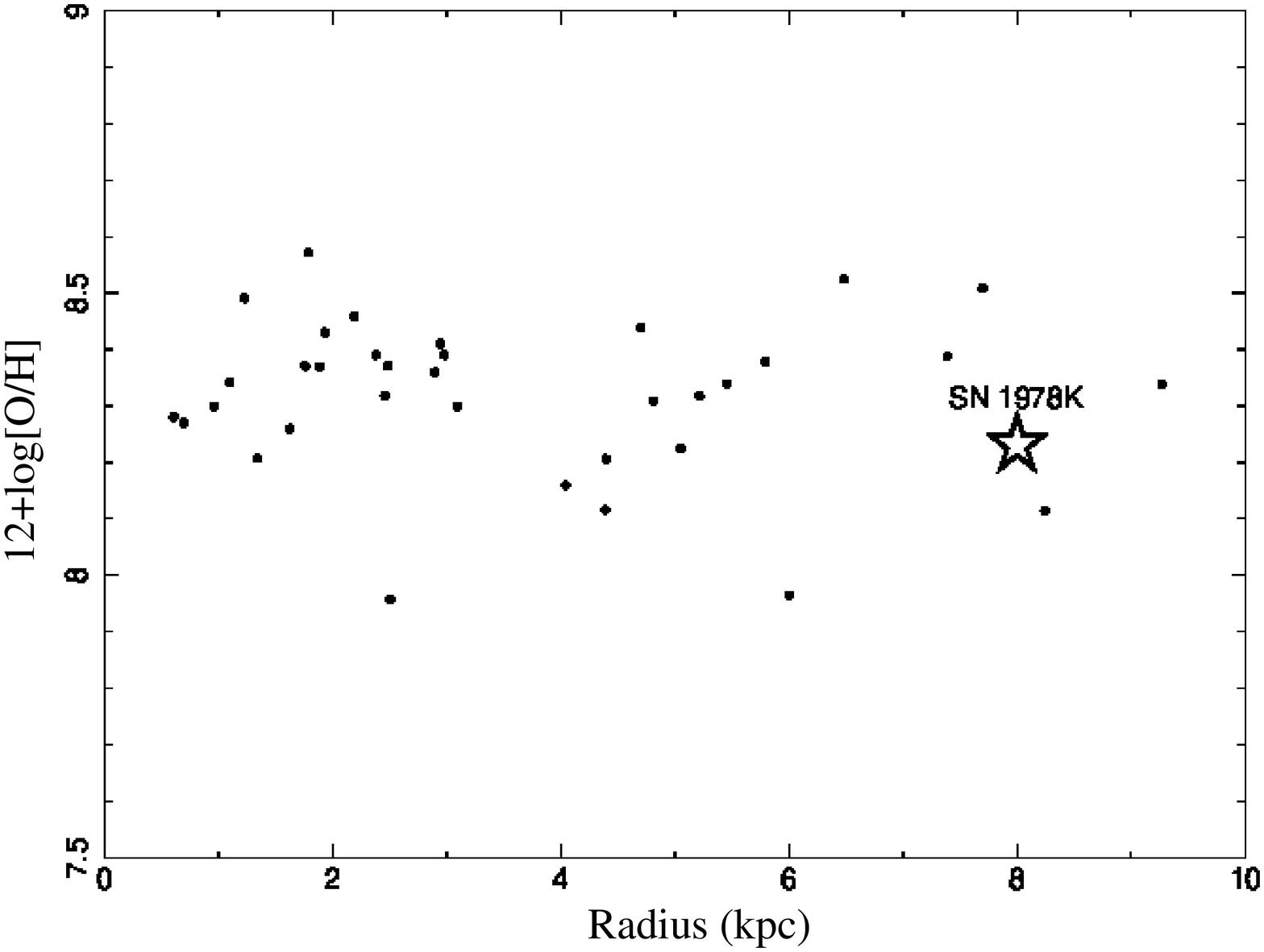} 
 \end{center}
\caption{NGC~1313 O/H abundance distribution based on observations of H II regions \citep{walsh1997,molla1999ngc1313metal}, with the SN~1978K data superimposed as a star.} \label{fig:ngc1313metal}
\end{figure}

\section{Discussion}

Based on a high-resolution RGS spectrum obtained over 11\,yr with a summed exposure time of 720\,ks (see Table~\ref{tab:obs_info}), we finally discovered solid evidence for X-ray line emission from N, O, Fe, and Ne for the first time from SN~1978K.  We also investigated a stacked MOS spectrum with a total exposure time of 900\,ks to support the RGS which is not sensitive to the higher energy above $\sim$2\,keV.  We confirmed no detection of Si lines from the stacked MOS spectrum, which is consistent with the result by \citet{lenz2007silver}. 

The metal abundances obtained are generally low, which clearly shows that the origin of the X-ray--emitting plasma is not metal-rich SN ejecta, but rather shock-heated interstellar and/or circumstellar matter.  We checked that our measured O/H abundance is consistent with that expected for a normal interstellar medium in NGC~1313, as shown in Figure~\ref{fig:ngc1313metal} which exhibits the O/H abundance distribution measured from H II regions in NGC~1313 \citep{walsh1997,molla1999ngc1313metal}, along with the data point for SN~1978K as a star mark.  In addition, the high N abundance (N/O $\sim 12$ solar) and the high density (10$^{5-6}$\,cm$^{-3}$: e.g., \cite{chu1999detection,kuncarayakti2016evolving}) led us to conclude that the plasma in SN~1978K originates from the CSM, i.e., the outer layer erupted from the progenitor star by stellar winds before the explosion.  

This fact assures that X-ray data can be used to estimate the progenitor's mass-loss rate (which was assumed a priori in previous X-ray analyses).  We inferred the mass-loss rate by comparing observed X-ray luminosities in Figure~\ref{fig:flux} with an analytical model, i.e., Equation [3.18] in \citet{fransson1994circumstellar}.  By integrating  the equation from 0.5\,keV to 2.0\,keV, it can be rewritten as $L_{\rm{cs}} \sim 6.15 \times 10^{38} (3-s)^{-1} T_{9}^{0.16} \times e^{-0.0116/T_{9}}  (\dot{M}_{-4}/V_{\rm{w2}})^{2}(V_{\rm{sh}}/10^{4}\ \rm{km\ s^{-1}})^{3-2s} \times (\it{t}/\rm{11.57}\ \rm{days})^{3-2s}\ \rm{erg\ s^{-1}}$, where a plasma temperature $T_{9} = (T/10^{9}\ \rm{K})$, a mass-loss rate $\dot{M}_{-4} = (\dot{M}/10^{-4}\ \rm{M_{\odot}\ yr^{-1}})$, and a wind velocity $V_{\rm{w2}} = (V_{\rm{w}}/100 \ \rm{km\ s^{-1}})$. The X-ray spectra required two-temperature components, i.e., low temperature ($T_{\rm{l}}$) and high temperature ($T_{\rm{h}}$), and we set $T_{\rm{l}} = 7.31 \times 10^{6}\  \rm{K}$ and $T_{\rm{h}} = 3.18 \times 10^{7}\ \rm{K}$ with their absorption-corrected luminosities $L_{\rm{l}} = 8.72 \times 10^{38}\ \rm{erg\ s^{-1}}$ and $L_{\rm{h}} = 3.82 \times 10^{38}\ \rm{erg\ s^{-1}}$, respectively.  Both of the two components must come from the CSM, because the elemental abundances are tied together between the two components in our analysis.  More specifically, we speculate that the low- and high-temperature components are due to a slower shock transmitting into the dense CSM and a reflected shock propagating back into the shocked CSM, respectively, as with the case of SN~1987A (e.g., \cite{zhekov2006}). In this interpretation, the constant luminosity for the low-temperature component indicates the density slope of the CSM to be s = 1.5.  We adopted $V_{\rm{sh}} = 1500\ \rm{km\ s^{-1}}$, the value estimated in \citet{kuncarayakti2016evolving} for the forward shock velocity.  The wind velocity, $V_{\rm{w}} = 100 \ \rm{km\ s^{-1}}$, was based on \citet{chu1999detection}.  We also adopted $t = 12000\ \rm{days}$, which is the mean days after explosion for the data sets analyzed in this work. Substituting these values into above $L_{\rm{x}}$--$\dot{M}$ relation, we obtained the mass-loss rate to be $\dot{M_{\rm{l}}} = 19.5 \times 10^{-4}\ \rm{M_{\odot}\ yr^{-1}}$ and $\dot{M_{\rm{h}}} = 6.2 \times 10^{-4}\ \rm{M_{\odot}\ yr^{-1}}$ for the low- and high-temperature components, respectively.  This result is well within the range of mass-loss rates estimated in preceeding works (e.g., \cite{ryder1993sn,chugai1995massloss,schlegel1999physical,tanaka2012infrared,kuncarayakti2016evolving}). According to \citet{kiewe2011caltech}, such a high mass-loss rate can be best explained by LBV stars whose mass-loss rates are measured to be $\dot{M} = 10^{-4} - 10\ \rm{M_{\odot}\ yr^{-1}}$.

Meanwhile, the abundance pattern of N/O $\sim$ 12 (N/O)$_\odot$ and O/H $\sim$ 0.2 (O/H)$_\odot$ measured for SN~1978K is indicative of lightly CNO-processed stellar material.  This is reminiscent of H-rich envelopes of red supergiants (RSGs), rather than CNO-equilibrium abundances with N/O = 26 (in number) and a significant depletion of O, as expected for LBV/WR stars.  Because the CSM we see in SN~1978K must have been ejected $\lesssim$300\,yr\,($V_{\rm sh}/600$\,km\,s$^{-1}$)\,($V_{\rm w}/100$\,km\,s$^{-1}$)$^{-1}$\,($t$/40\,yr) before the explosion (e.g., \cite{ryder2016,kuncarayakti2016evolving}), their abundance would reflect that of the surface of the final fate of a progenitor star.  Therefore, the abundance pattern suggests that the progenitor star of SN~1978K did not evolve into a LBV phase at the time of explosion.  

We note that optical spectroscopy revealed narrow, unshocked H$\alpha$ and [N II] emission \citep{chu1999detection}, with a [N II]/H$\alpha$ ratio of 0.8--1.3, which is an order of magnitude higher than what is expected for an interstellar HII region, and is probably close to our X-ray measurements.  \citet{gruendl2002} confirmed this result, and gave an upper limit of $\sim$2.2\,pc for the size of the CSM nebula.  Such a high [N II]/H$\alpha$ ratio and the small size are similar to ejecta nebulae around LBV stars.  However, the nebulae around LBV/WR stars and the HII region around SN~1978K are thought to have been ejected during RSG phases, i.e., some 10$^4$\,yr before the LBV phases.  Thus, this does not conflict with our interpretation that the progenitor star of SN~1978K did not reach the LBV/WR phase.  

In summary, the high mass-loss rate suggests a massive LBV-like progenitor, whereas the abundances of N and O indicate a relatively low-mass progenitor.  We here consider three possibilities to solve this contradiction.  

The first possibility is an electron-capture SN (ECSN) from a super asymptotic giant branch (SAGB) star with masses of 8--10\,M$_\odot$.  The N/O abundances at surfaces of SAGB stars are expected to be around 10 solar, independent of the initial metallicity \citep{doherty2014}, which is consistent with our measurements.  Other remarkable explosion properties that are thought to be ECSNe are (1) low velocities of 500--1000\,km\,s$^{-1}$, (2) spectra similar to SNe IIn, and (3) low peak luminosities of -10 to -15 mag, as summarized in \citet{adams2016}.  The former two characteristics are indeed consistent with those of SN~1978K.  As for the peak luminosity, it is difficult to assess the maximum brightness of SN~1978K, because there are no observations that cover the time period around the peak luminosity.  But, there is a hint that SN~1978K was an under-luminous, low-energy explosion; \citet{kuncarayakti2016evolving} estimated an explosion energy to be 0.1--2 $\times$ 10$^{51}$\,(M$_{\rm ejecta}$/10\,M$_\odot$)$^{0.5}$\,ergs, based on the CSM density combined with an SNR evolution model by \citet{chevalier1994}.  If the progenitor star is indeed an SAGB star, then we expect the ejecta mass to be a few solar masses, so that the explosion energy would decrease half, suggesting a low-energy explosion for SN~1978K.  Finally, stellar population analysis also indicates a relatively low-mass progenitor, M$_{\rm ZAMS}=8.8\pm$0.2\,M$_\odot$ \citep{Williams2018mass}, which is exactly what we expect for an SAGB star.  We should keep in mind, however, that there are two weaknesses of the SAGB scenario.  First, it is difficult to make the large mass loss of the order of $10^{-3}$\,M$_\odot$\ $\rm{yr}^{-1}$ over 1000\,yr long (as observed), even during the thermal-pulse phases \citep{doherty2015}.  Second, the progenitor star detected with the absolute magnitude of M$_{\rm B} \sim -6$ (corresponding to log(L/L$_\odot$) $\sim$ 4.6) may be too bright, given the possible dusty environment inferred from the high mass-loss rate and bright IR emission from SN~1978K.  In fact, the progenitor star detected for a proto-typical ECSN candidate, SN~2008S, was heavily obscured and undetected in the optical band \citep{prieto2008}.

The second possibility is an Fe core-collapse SN of a relatively low-mass ($\sim$10\,M$_\odot$) red supergiant star, whose properties are similar to those for ECSNe from a theoretical point of view.  One strength with this scenario would be that the RSG progenitor could produce the high mass-loss rate of the order of 10$^{-3}$\,M$_\odot$\,yr$^{-1}$ \citep{vanloon1999}, although there is no convincing theory to explain such high mass-loss rates (but see below for ideas).  To estimate the surface N/O ratio, we simulated a stellar evolution for an initial mass of M$_{\rm ZAMS} =$ 8\,M$_\odot$ and a metallicity of Z $\sim$ 0.3\,Z$_\odot$ (cf.\ Z = 0.24\,Z$_\odot$ was obtained for the CSM in SN~1978K; see Table~\ref{tab:spec_fitpara}), using HOSHI code with the same stellar parameters of Set L$_{\rm{A}}$ in \citet{yoshida2019} (see also \cite{takahashi2016,takahashi2018}).  As a result, we found that the surface abundance of N/O becomes 4, 6, and 10 times the solar value for initial rotation velocities of 0\,km\,s$^{-1}$, 200\,km\,s$^{-1}$, and 262\,km\,s$^{-1}$, respectively.  Therefore, given that the average rotation velocity for massive stars is 190\,km\,s$^{-1}$ (p.686, Figure~27.1 in \cite{maeder2009}; see also Figure~6 in \cite{huang2006}), the N/O $\sim$12 solar measured for SN~1978K requires a relatively fast rotation.

The third possibility is an Fe core-collapse SN of a relatively massive (20--25\,M$_\odot$) RSG star.  We simulated a stellar evolution for M$_{\rm ZAMS} =$ 25\,M$_\odot$ and Z $\sim$ 0.3\,Z$_\odot$, and found that the surface N/O ratio suddenly increased up to $\sim$15 times the solar value at $\sim$2$\times$10$^5$\,yr before the explosion, when the star evolves from a yellow supergiant to a red supergiant.  Thus, if large mass losses in the RSG phase take place, then the observed high N/O ratio and high mass-loss rate could be explained by this model.  Although our current model cannot explain the high mass-loss rate of the order of 10$^{-3}$\,M$_\odot$\ yr$^{-1}$, there are possibilities that efficient mass losses are caused by the decrease of the gravity during the RSG phase.  For instance, it is suggested that pulsation-driven superwinds would induce such a high mass loss in a RSG phase \citep{yoon2010}.  Also, \citet{shiode2014} argued that during and after core Ne burning, internal gravity waves excited by core convection can tap into the core fusion power and transport a super-Eddington energy flux out to the stellar envelope, potentially unbinding $\sim$1\,M$_\odot$ of material.  As for the other properties of SN~1978K, the low explosion velocity (or low explosion energy) is consistent with those for low-luminosity Type IIP SNe such as SN~1997D, for which recent works seem to support high-mass ($\gtrsim$25\,M$_\odot$) or intermediate-mass (10--15\,M$_\odot$) progenitor scenario (e.g., \cite{turatto1998,zampieri2003,spiro2014}), although a low-mass ($\sim$10\,M$_\odot$) progenitor scenario is not yet fully ruled out \citep{chugai2000}.  Also, in the RSG scenario, the progenitor star's brightness of log(L/L$_\odot$) $\sim$ 4.6 could agree with those for typical RSGs (log(L/L$_\odot$) $\sim$ 5.5) by taking into account the dust extinction with an optical depth of a few.  The presence of dust extinction is at least qualitatively more realistic than little or no dust extinction required for the SAGB scenario as described above.  We however point out two problems with this scenario.  First, an RSG progenitor should produce a Type IIP SN rather than IIn, conflicting with the IIn identification for SN~1978K.  Second, there are no promising physical reason to explain the high mass-loss rate of the order of 10$^{-3}$\,M$_\odot$\ yr$^{-1}$ (but see above for some ideas).  In this respect, further theoretical studies are eagerly awaited.

Finally, we point out yet another possibility that SN~1978K was actually an SN impostor such as SNe~1961V \citep{vandyk2012} and 2009ip \citep{margutti2014}.  The peculiar properties of SN~1978K ($\dot{M} \sim 10^{-3}$\,M$_\odot$\ $\rm{yr}^{-1}$, N/O $\sim$ 10 solar, expansion velocity $\sim$ 500\,km\,s$^{-1}$) are all shared with SN impostors.  Continued follow-up observations of this long-lasting SN are important to reveal a massive progenitor (if any) that survived the giant eruption in 1978, and to study the long-term mass-loss history with extreme mass-loss rates.

\section{Conclusions}

We analyzed Type IIn SN~1978K's X-ray spectra obtained with \textit{XMM-Newton}'s RGS.  By stacking the data, we created a long-exposure spectrum, from which we detected a number of emission lines from N, O, Ne, and Fe for the first time from SN~1978K.  Notably, we found a high abundance of N, N/O $\sim 12$ solar, and a low metallicity, Z $\sim$ 0.24\,Z$_\odot$.  Such an abundance pattern suggests that the blastwave of SN~1978K is interacting with a lightly CNO-processed CSM.  Combined with the low expansion velocity and SN~IIn-like optical spectra, led us to consider that the origin of SN~1978K was either an ECSN from a SAGB star or a weak core-collapse explosion from a RSG star with a mass of either $\sim$10\,M$_\odot$ or $\sim$25\,M$_\odot$.  From the X-ray luminosity, we obtained a mass-loss rate of the progenitor star to be about $10^{-3}\,\rm{M_{\odot}\ yr^{-1}}$.  The high luminosity of SN~1978K over 40\,yr after the explosion suggests that such a high-mass loss rate lasted for more than 1000\,yr before the explosion.  Such extreme mass losses remain a question to the study of massive stars' evolution.

\begin{ack}
We would like to thank Drs.\ Makoto S. Tashiro, Yukikatsu Terada, Kosuke Sato, Liyi Gu, and Takaya Nozawa for a number of constructive comments that improved the quality of this paper.  This work was supported by the Japan Society for the Promotion of Science KAKENHI grant numbers JP17H02864 (SK) and 17H01130 (HU).  This work was partly supported by Leading Initiative for Excellent Young Researchers, MEXT, Japan.

%(NOT after the Appendix.)
\end{ack}

%\appendix 
%\section*{Case of single paragraph}

%\section{Case of two or more paragraphs}

%\section{Case of two or more paragraphs}

%%%
% See the manual for the detail.
%%%

%\bibitem[ et al.()]{}
%
%\bibliography{cite} 
%\bibliographystyle{plain}

\end{document}